\documentclass[useAMS,usenatbib]{mn2e}

\usepackage{graphicx}
\newcommand{\afe}{{[\alpha/\mathrm{Fe}]}}
\newcommand{\feh}{{[\mathrm{Fe}/\mathrm{H}]}}
\newcommand{\msun}{{\mathrm{M}_\odot}}
\newcommand{\rsun}{{\mathrm{R}_\odot}}
\newcommand{\msunpy}{{\mathrm{M}_\odot}\,\mathrm{yr}^{-1}}
\newcommand{\ace}{{\alpha_{\mathrm{CE}}}}
\newcommand{\lce}{{\lambda_{\mathrm{CE}}}}
\newcommand{\Mch}{{\mathrm{M}_\mathrm{Ch}}}

\title[Impact of binary star yields on spectra]{The impact of binary star yields on the spectra of galaxies}
\author[Sansom, Izzard \& Ocvirk]{A. E. Sansom$^{1}$\thanks{E-mail:
    AESansom@uclan.ac.uk (AES)}, R.G. Izzard$^{2,3}$, P. Ocvirk$^{1,4}$ \\
$^{1}$Jeremiah Horrocks Institute for Astrophysics and Supercomputing, University of Central Lancashire, Preston, Lancashire, PR1 2HE, UK\\
$^{2}$Sterrenkundig Institute, Universiteit Utrecht, PO Box 80000,
3508 TA, Utrecht, The Netherlands\\
$^{3}$Université Libre de Bruxelles, Boulevard du Triomphe, B-1050 Brussels, Belgium\\
$^{4}$Astrophysikalisches Institut Potsdam, An der Sternwarte 16, D-14482 Potsdam, Germany}

\begin{document}

\date{}


\maketitle

\label{firstpage}

\begin{abstract}
One of the complexities in modelling integrated spectra of stellar populations 
is the effect of interacting binary stars besides type Ia supernovae (SNeIa). 
These include common envelope systems, CVs, novae, and are usually ignored 
in models predicting the chemistry and spectral absorption line strengths 
in galaxies. 
In this paper predictions of chemical yields from populations of 
single and binary stars are incorporated into a galactic chemical evolution 
model to explore the significance of the effects of these other binary yields. 
Effects on spectral line strengths from different progenitor channels of 
SNeIa are also explored. Small systematic effects are found when the yields 
from binaries, other than SNeIa, are included, for a given star formation 
history. These effects are, at present, within the observational 
uncertainties on 
the line strengths. More serious differences can arise in considering 
different types of SNIa models, their rates and contributions.

\end{abstract}


\section{Introduction}
\label{section-introduction}
Spectral absorption line strengths of galaxies can be used to probe  
their star formation histories (SFHs). These techniques are becoming 
increasingly refined and better understood in terms of breaking 
underlying degeneracies in age and metallicity (\citealp{Worthey1994}; 
\citealp{Trager2000}; \citealp{Proctor2002} - hereafter PS02; 
\citealp{Trager2005}; \citealp{Kaviraj2007}; \citealp{Smith2008}) 
and accurate interpretation of integrated light observations 
(e.g. \citealp{Proctor2004}; \citealp*{Thomas2004}; 
\citealp{Lee2007}; \citealp{Serra2007}). 
Uncertainties in stellar population models 
limit the effectiveness of these studies. One outstanding 
question that needs to be addressed is: how are the observed spectral 
line strengths in integrated stellar populations affected by abundance 
changes due to interacting binary stars? Most modellers of composite 
stellar populations take into account the element yields from SNeIa 
explosions from interacting binary stars 
(normally the Chandrasekhar mass, W7 model of 
\citealp*{Nomoto1984}) however, yields from other types of 
interacting binary star processes, such as various common envelope 
binaries that do not necessarily become SNeIa, are seldom taken into account.

In most studies, unresolved stellar populations in galaxies are 
characterised by their luminosity weighted average properties, by fitting 
observed spectral line-strengths to predictions from single age, single
metallicity stellar population (SSP) models (e.g. PS02, \citealp{Trager2005}).
The system of line strengths most widely used is that of Lick indices
(\citealp{Worthey1994a}; \citealp{Worthey1997}) based on 
observations of stars taken at the Lick observatory. This is the system of 
line-strength definitions and calibration star data that is used in this 
paper, to link effectively to observable properties of galaxy spectra.
 
In some cases more complex SFH fitting, to line strengths or colours, 
is attempted using e.g. multiple
SSPs (e.g. \citealp{Heavens2004}, \citealp{Rossa2006}, \citealp{Serra2007}) 
or an exponentially decaying SFH (\citealp{Fritze1994}; 
\citealp{Li2004}; \citealp{Ganda2007}) sometimes with optional starbursts 
(\citealp{Temporin2006}). Empirical fitting of too 
many SSPs is undermined by the degeneracy 
between parameters. Optical spectra of old, metal-poor SSPs can look 
remarkably similar to those of younger, more metal-rich SSPs, because 
both increasing age and increasing metallicity depreciate the blue continuum. 
Additional constraints are provided by making the 
SFH self consistent with the feedback into the inter-stellar medium 
(ISM) from which stars are made. This is the approach of galactic 
chemical evolution (GCE) modelling, where elemental abundances are 
followed for a given SFH, in order to model composite stellar populations.

Single SSP fitting is more robust than trying to fit the full SFH, 
and SSP fitting can be carried out for lower signal-to-noise data 
($\mathrm{S/N}>15/\mathrm{\AA}$). However an SSP is a poor representation
of the SFH of a real galaxy. On the other hand, GCE modelling may lead 
to a more realistic determination of the SFH, but 
requires higher signal-to-noise data ($\mathrm{S/N}>30 \mathrm{\AA}$), 
depending on the complexity of the underlying SFH. Therefore the two 
techniques provide 
complementary approaches to determining when stars formed in galaxies.
 
Full spectrum fitting seems a promising alternative to Lick index 
fitting \citep{Ocvirk2006} but there is currently no model that predicts the 
full (visible) spectrum of stellar populations with a range of non-solar 
abundance ratios, even though progress is being made with this 
(\citealp{Coelho2007}; \citealp{Lee2009}). In this present work we limit 
ourselves to fitting Lick indices.

\citet{Zhang2005} calculated the effects on luminosity weighted age
and metallicity estimates, derived from spectral line strengths and
colours, when binary interactions are included in the calculations 
of stellar evolution. They considered the extreme case of 100\% stars 
in binaries. Their models were based on the rapid binary star evolution 
code of \citet*{Hurley2002}, which predicts physical aspects of binary 
star components, such as their mass, size, separation, luminosity and 
temperature etc. with time. They took into account the 
evolution and effects on spectral characteristics of the above physical 
effects of binary stars in a population, and showed what this meant for 
line strengths in SSPs of different ages and metallicities. 
Although they concluded that it is important to model binary stars, 
their results actually show fairly small changes in most line strengths 
because of binary evolution, with changes in line strengths of up to a 
few times the typical state-of-the-art observational errors, as 
discussed in Section~\ref{obs-errors} of this paper. For example, 
changes in $\mathrm{H}\beta$ were up to $+0.37 \mathrm{\AA}$ for 
populations with minus without binary interactions.
\citet{Li2008} find similar results. 
However, they did not model the yields from binary stars, 
nor did they consider non-solar element abundance ratios in their SSPs.

\citet{DeDonder2004} combined initial properties of stars with 
stellar evolutionary models (theirs and others) to generate the yields 
from stellar populations including binary types.
They found that including the effects of binaries (apart from SNeIa) 
leads to small changes in the predicted yields. Other uncertainties in 
stellar evolution (e.g. mass-loss rates, massive-star remnant masses etc.) 
make at least as much difference to the chemical yields.

In this paper we take a complementary approach and consider the effect of
incorporating yields from single and binary star populations into a GCE model
in order to determine the effect of chemistry on absorption line strengths.
We apply the code of \citet{Izzard2004} and \citet{Izzard2006} 
to calculate the yields of both single- and binary-stellar populations of 
different metallicities as a function of age. 
In that work, as expected, the main difference between single- 
and binary-star populations is the effect of SNeIa, but yields from giant
stars are also reduced in binary populations. This is because when a star
becomes a giant it is most likely to interact with a close companion and
lose mass through non-conservative Roche-lobe overflow and common-envelope 
evolution. The yields of elements produced in asymptotic giant branch (AGB) 
stars e.g. carbon,
nitrogen and $s$-process elements such as barium, are reduced by up to $20\%$
compared to single stars.

We combine the single- and binary-star yields with a galactic chemical 
evolution model which
parameterises star formation histories with six variable parameters. 
These are tuned to simulate a range of galaxy types (PS02).

Unfortunately, a major uncertainty in our yield calculation, and hence in 
our GCE models, arises from the calculation of yields from type Ia supernovae. 
While our binary population synthesis model predicts a certain Ia rate based 
on the algorithm of \citet*{Hurley2002}, there are several progenitor channels,
all of which are seriously affected by model parameters (e.g. 
the common-envelope ejection efficiency). Worse still, it is not known which
progenitor channels are truly responsible for SNeIa.

In order to consider quantitatively this uncertainty, we have split our binary 
yield set into two parts: SNeIa and ``other binary yields''. The SNeIa set 
is then subdivided into two channels: 
1) Sub-$\Mch$ explosions, which in our model arise from edge-lit detonations 
(ELDs), where $\Mch \sim 1.4 \msun$ is the Chandrasekhar mass. These explode 
when a CO white dwarf (COWD) accretes more than $0.15 \msun$ of helium-rich 
material.
2) $\Mch$ explosions arise from the accretion of material onto a COWD until 
it reaches a mass of $\Mch$. This set includes COWD-COWD mergers.

The ``other binary yields'' set, which includes yields due to all other 
processes in binary stars such as non-conservative Roche-lobe overflow, 
common-envelope ejecta, novae, stellar winds etc., is then treated separately 
from the uncertainties in SNeIa yields.

There is a time delay between the formation of a binary system that could 
lead to a SNIa and the explosion itself. This time delay depends on the 
type of SNIa progenitor (e.g. \citealp{DeDonder2004}; \citealp{Greggio2005}) 
and this affects the timescale of enrichment of the ISM with iron in galaxies. 
Thus it is important to study the effects of different SNeIa progenitors and 
the parameters associated with their production rates. 

Previous authors have described how the various sources of enrichment 
are expected to contribute to the yields with time in a stellar population
(e.g. \citealp{Worthey1998}; \citealp{Ballero2007}). Briefly, type II SNae 
from massive stars will contribute a broad range of metals including the 
important $alpha$-capture elements (O,Ne,Mg,Si,S.Ar,Ca etc.) on short 
timescales ($\le5\times 10^7$ years, approximately the pre-supernova 
lifetime of an $8\msun$ star); whereas SNeIa from interacting binary star 
channels contribute mainly iron-peak elements (Cr,Fe,Ni), over longer 
timescales (e.g. \citealp{Mannucci2006}). Intermediate mass single stars 
recycle H and He and also contribute to He, C, N and O yields.
Intermediate mass stars in binary systems other than SNIa also contribute
small amounts of other heavy elements (e.g. \citealp{DeDonder2004}; 
\citealp{Izzard2006}). These different sources are all included in our 
GCE models.

This paper addresses the question of how these yields affect spectral 
line strengths in self-consistent GCE models, covering a variety of 
star formation histories. Variations of abundances and abundance ratios 
are taken into account through the SSP line-strength models used. 
Section~\ref{section-models} describes our models of 
binary-star yields and our GCE models.
Section~\ref{section-effect-of-binaries} describes how these combined
models are used to test the effects of including other binary yields, 
besides SNeIa, on line strengths; 
Section~\ref{section-different-sneia} explores some effects of 
different SNIa models and common envelope parameters on line strengths;
some discussion is given in Section~\ref{section-discussion} 
and Section~\ref{section-conclusions} summarises our results.

\section{Models}
\label{section-models}
In this section we briefly describe our models.
We begin with our single- and binary-star stellar yield calculations 
and follow with a description of their implementation into our galactic 
chemical evolution code.

\subsection {Yields from single and binary stars}

\label{yield-calculations}

Our yields are calculated for single and binary stars according to the
rapid binary evolution and nucleosynthesis models of 
\citet{Izzard2004} and \citet{Izzard2006}. Stellar evolution is
based on the algorithm of \citet{Hurley2002} which uses analytic fits
to stellar properties such as luminosity, radius and core mass to follow
stellar evolution for stars of masses up to $100 \msun$ 
from the main sequence to the end of evolution, either as a white dwarf or
in a supernova explosion. This is coupled to
a binary interaction scheme which models tidal interaction, mass transfer 
by winds and Roche-lobe overflow, mass accretion, common envelope evolution, 
novae etc.

The nucleosynthesis routine runs in parallel to the stellar evolution 
algorithm and follows the surface abundances of stars during all phases 
of evolution. It includes the following:
 
\begin{itemize}
\item A synthetic thermally pulsing asymptotic giant branch model 
which mimics the behaviour of the detailed
models of \citet{Karakas2002} and \citet{Karakas2007}, including third dredge up and hot-bottom burning by the CNO, NeNa and MgAl cycles.
\item Surface abundances, and hence stellar wind yields, from massive stars according to \citet{Dray2003a}, \citet{Dray2003b} and Stancliffe (private communication).
\item Core-collapse supernovae according to the calculations of \citet{Chieffi2004}.
\item SNIa yields from sub-Chandrasekhar mass (sub-$\Mch$) 
and Chandrasekhar mass ($\Mch$)
explosions (yields from \citealp{Livne1995} and the DD2 model of
\citealp{Iwamoto1999} respectively).
\item Nova yields of \citet{Jose1998}.
\item Stellar wind collision, accretion and time-dependent 
thermohaline mixing into secondary stars.
\end{itemize}

Many parameters are associated with both our single- and binary-star 
models, these are discussed in some detail in \citet{Izzard2006}.
We make no changes from the standard assumptions used there, except
that orbital energy transfer parameter $\ace$ is set to $3$ as 
suggested by \citet{Hurley2002}
in order to maximise the effect of SNeIa. 
The common envelope structure parameter $\lce$ is fitted to the data of 
\citet{Dewi2000} and \citet{Tauris2001} although we optionally allow
for a fixed value of $0.5$ 
(see Section~\ref{effects-of-different-sneia}).

This constitutes our ``baseline model'' referred to in later sections.
We note that the rate (and hence yield) of $\Mch$ SNeIa is considerably
less than typically used in GCE models, but we make up for this with
copious numbers of $\mathrm{sub-}\Mch$ supernovae. A detailed
study of the rates and yields of SNeIa is beyond the scope of this
paper but a preliminary exploration is carried out in 
Section~\ref{section-different-sneia}.

Our yield sets are calculated by integrating yields for a population of 
stars using the \citet*{Kroupa1993} initial mass function (IMF) with masses 
between $0.1$ and $80 \msun$ for single stars and binary primary stars, a
binary secondary mass distribution which is flat in mass ratio $q$ such that 
all secondary masses (less than the primary mass i.e. $q<1$) are equally 
likely, and a binary separation distribution which is flat in 
$\log$-separation (i.e. distributed as $1/a$ where $a$ is the binary 
separation) between $3$ and $10^4 \rsun$. The yields were generated in
$1 \mathrm{Myr}$ bins for a maximum time baseline of $14 \mathrm{Gyr}$.
It proved difficult to temporally resolve yield output at late times
when using our standard logarithmic grid for the primary mass.
To resolve this we applied an algorithm which chooses the primary
masses as a function of the main-sequence lifetime
such that yield rate at late times is time-resolved and smooth.

We split our yield calculations into three sets:

\begin{itemize}
\item Single stars ($s$),
\item Binary stars, excluding SNeIa ($b$), and
\item SNeIa (of all types) ($SNIa$).
\end{itemize}

These contributions to yields are combined in different 
proportions in our GCE models (see Sections~\ref{GCE-models} and 
\ref{section-effect-of-binaries}).

\subsection {Galactic chemical evolution models}
\label{GCE-models}

Our GCE model self-consistently follows the mass and 
chemistry of both stars and the ISM for a given SFH (see PS02 for a 
detailed description) which is described by a 
simple but flexible parameterisation. 
A Salpeter IMF is assumed for the stars, as available for the SSPs.
A Schmidt law is assumed for the star formation rate: $\mathrm{SFR}=Cg$, 
where $g$ is the gas mass within a fixed volume and $C$ is the efficiency 
of star formation. Sixteen of the most abundant elements are followed 
individually in our GCE model and are then combined into groups when 
linked to SSP models. The chemistry and mass of stars thus generated 
as a function of time are converted into observables through the use of SSPs. 
We use line strength predictions from the SSP models of \citet*{Thomas2004} 
- hereafter TMK04. These allow not only for age and metal mass fraction 
($Z$) variations in stellar populations but also for dependencies of line 
strengths on element abundance ratios, characterised by the ratio of 
$\alpha$-capture to iron-peak elements ($\afe$
\footnote{Where $\afe=\log(\alpha/\mathrm{Fe})-\log(\alpha_\odot/\mathrm{Fe}_\odot)$
, with $\alpha$ and Fe being the gas mass fractions of $\alpha$-capture 
and Fe-peak elements from which the stars are made. See TMK04 and references 
therein for details of the elements included.} 
ratio). Following TMK04 we combine the abundant $alpha$-capture elements 
O,Ne,Mg,Si,S,Ar,Ca plus N and Na as the $\alpha$-elements group of 
elements in order to predict [$\alpha$/Fe] for the SFHs modelled. Carbon 
remains separate from the $\alpha$-elements in the models of TMK04.

The parameters that describe the SFHs of our GCE models are listed in 
Table~\ref{SFH-table} and a schematic illustration is given in Fig.~A1 
of Appendix~\ref{GCE-parameters}. 
These allow for a wide range of possible SFHs, covering early star 
formation (SF), described by the SF efficiency ($\mathrm{C}_0$) and gas inflow 
rate ($\mathrm{F}_0$), plus an optional, more recent SF event. There are 
parameters for the onset time ($\mathrm{T}_1$), duration ($\mathrm{D}_1$) 
and star formation efficiency ($\mathrm{C}_1$) of this later event, as well 
as the gas inflow rate ($\mathrm{F}_1$) during this more recent SF event. 
The SF and gas flow are switched off after the merger event because 
elliptical galaxies show little evidence for ongoing star formation or 
massive gas flows. This star formation shutoff (often referred to 
as ``quenching'') mimics the fact that the remaining gas has been made 
unavailable for star formation through some process, either Active 
Galactic Nucleus feedback \citep{Bower2006,Hopkins2007}, or heating 
through accretion shocks \citep{DB2006,Ocvirk2008}. 
Units for these parameters are also given in Table~\ref{SFH-table}.

\begin{table*}
\begin{minipage}{150mm}
\begin{tabular}{cccccccc}
\hline

{Scenario}&{${\rm{C}}_{0}$} & {${\rm{F}}_{0}$} & {${\rm{C}}_{1}$} 
& {${\rm{T}}_{1}$} & {${\rm{F}}_{1}$} & {${\rm{D}}_{1}$} 
& {Comment} \cr
           &{Gyr$^{-1}$} & {$\msun\,\mathrm{Gyr}^{-1}$} & {Gyr$^{-1}$} 
& {Gyr}       & {$\msun\,\mathrm{Gyr}^{-1}$} & {Gyr} & \cr

\hline
PC1&$4.0$& $4\times10^6$&$4.0$&$0.25$&$0$&$0.25$ & Early, rapid SF lasting $0.5\,\mathrm{Gyr}$ \cr 
PC2&$4.0$&$1.3\times10^6$&$4.0$&$0.75$&$0$&$0.75$ & Early, rapid SF lasting $1.5\,\mathrm{Gyr}$ \cr 
LB1&$4.0$&$0$&$4.0$&$8.7$&$2\times10^6$&$0.5$ & Early, rapid SF + burst $5\,\mathrm{Gyr}$ ago \cr 
LB2&$4.0$&$0$&$4.0$&$11.7$&$2\times10^6$&$0.5$ & Early, rapid SF + burst $2\,\mathrm{Gyr}$ ago \cr 
MMW&$0.04$&$0$&$4.0$&$11.7$&$2\times10^6$&$0.5$ & Early, slow SF + burst $2\,\mathrm{Gyr}$ ago \cr 
\hline
 \end{tabular}

\caption{Parameters that define the five star formation histories considered 
in this paper. 
\label{SFH-table}
The star formation efficiencies ${\rm{C}}_0$ and ${\rm{C}}_1$ are 
defined as in \citet{Sansom1998}, 
while ${\rm{F}}_{0}$ and $ {\rm{F}}_{1}$ are infall rates in 
$\msun/{\rm{Gyr}}$, with the initial gas mass being $10^6\,\msun$. 
${\rm{T}}_{1}$ and ${\rm{T}}_{2}$($={\rm{T}}_{1}+{\rm{D}}_{1}$) 
are the transition times in Gyr where the corresponding infall rates and 
star formation efficiencies apply (i.e. ${\rm{C}}_1$ and $ {\rm{F}}_{1}$ 
apply during ${\rm{D}}_{1}$). The motivation for
choosing these parameters is discussed in Appendix~\ref{GCE-parameters}. 
Each set of parameters 
corresponds to a different evolutionary scenario. From top to bottom: 
primordial collapse with infall and star formation shutting off early (PC1) 
and slightly later (PC2). The next two scenarios describe an early-type galaxy 
accreting a significant amount of gas (e.g. through a merger) at a late 
(LB1) or very late (LB2) age, 
and where this event triggers a burst of star formation that shuts off 0.5 Gyr 
after the encounter. The last scenario (MMW) represents a slowly forming, 
Milky Way-type 
disc accreting a large mass of gas that triggers intensive star formation 
at a late age. In each case the infalling gas doubles the mass. In these 
models the present age of Universe is assumed to be 13.7 Gyr.}

\end{minipage}
\end{table*}
 \vspace{0.3cm}

Our GCE model incorporates tables of yields from stellar models. For the 
present work, the yields used are from the single and binary star SSP 
models described in Section~\ref{yield-calculations}. 
The GCE code models the mass of stars and their element abundances for 
populations in each time step. Appropriate SSP model information 
(integrated line strengths and luminosities) is chosen for the relevant 
age, metallicity and $\afe$ ratio for stars at each time step. 
The luminosity weighted sum of SSPs then predicts the observable 
characteristics, i.e. the line strengths, of the integrated stellar 
population.

There are no SSP models of line strengths that incorporate {\it both} binary 
evolution effects {\it and} effects of non-solar abundance ratios. 
The former effect has been shown to be small by \citet{Zhang2005}.
The latter effect is important for modelling accurate line strengths 
in galaxies. Our SSP line-strength models are based on TMK04 
because they allow a variable $\afe$. However, TMK04 do not 
predict broadband colours, so our luminosity weightings 
(in \emph{B}, \emph{V} and \emph{I}) are taken from the SSP 
models of \citet[][see http://www.cida.ve/$\sim$bruzual/bc2003]{Bruzual2003}. 
Their luminosities for a 
Salpeter IMF are used, for consistency with the SSP line strength models,
and assuming Padova 1994 tracks. The filters used to compute 
the \emph{B}, \emph{V}, \emph{I} luminosities are respectively the 
Buser \emph{B3} filter, Buser \emph{V} filter and the Cousins \emph{I} filter.

Spectral line strengths are thus predicted for integrated populations  
containing yields from binary and/or single stars. Our GCE models 
have a fixed time step ($30 \mathrm{Myr}$) and line strengths are 
calculated at $13.7 \mathrm{Gyr}$ -- the approximate age of the Universe
from the WMAP 3 year results \citep{Spergel2007}.
 
To test the effects on yields and line-strength from binary stars, a range 
of simple SFHs were tested both with and without binary stars ($b$ above). 
The motivation for our choice of SFHs was to cover a range of representative 
histories approximating early-type (E,S0) galaxies (e.g. \citealp{Rakos2008}, 
\citealp{Reda2007}), including early (primordial) collapse models, early 
collapse followed by later star
formation to mimic later mergers (at 2 or 5 Gyrs ago) and finally a model
approximating the rapid merger 2 Gyr ago of two spiral galaxies like the 
Milky Way in their pre-merger SFR, accompanied by a merger induced starburst.
An example of a system which has undergone a major merger in the last 2 Gyr
is NGC 2865 (\citealp{Hau1999}).
The SFH models explored here do not attempt to model recent SF or late-type 
galaxies. Instead we concentrate on modelling the effects of binary star 
yields on galaxies that are early-types now, for which the question of binary 
star effects on absorption line strengths is more pressing due to the slow 
changes of spectral features with stellar population parameters.
These SFHs are listed and described in Table~\ref{SFH-table} (and
shown later, in Fig.~\ref{Fig-3}). The binary 
fraction is uncertain (e.g. \citealp{Converse2008} and references therein). 
Here, for simplicity, we assume $50\%$ mass fraction in single stars and 
$50\%$ in binary stars. First, the effects of yields from 
binary stars other than SNeIa explosions were tested 
(Section~\ref{section-effect-of-binaries}), 
using normalisations based on recent observations 
(described in Section~\ref{obs-errors}).
Then the effects of different assumptions about SNeIa progenitors and common 
envelope parameters were tested 
(Section~\ref{effects-of-different-sneia}).

\section{Effects of binary yields other than SNeIa}
\label{section-effect-of-binaries}
Canonical galactic chemical evolution models treat all stars as single
stars and implement binary effects with ad-hoc 
prescriptions for type Ia supernovae and, rarely, yields from 
novae\footnote{Nova yields are probably only relevant for minor isotopes 
such as $^{13}\mathrm{C}$.}. Our calculation differs somewhat in that we 
calculate SNIa and other binary star yields (described in 
Section~\ref{yield-calculations}) directly
from our binary population simulations, so an ad-hoc model is not required.
Therefore we implement true binary yields, including the SNIa yields, as 
calculated in section~\ref{yield-calculations}.
We consider a binary fraction of $50\%$ by mass\footnote{Equivalent to 
$\sim 40\%$ binary fraction by number of systems assuming our default 
initial distributions of binary parameters.} such that our yield sets are: 

\begin{eqnarray}
\frac{1}{2}s+\frac{1}{2}b+\frac{1}{2}\mathit{SNIa} & \mathrm{(hereafter\, 
population\, B)} \label{Eq-1} & \mathrm{and}\\
s+\frac{1}{2}\mathit{SNIa,} & \mathrm{(hereafter\, population\, S)} \label{Eq-2}
\end{eqnarray}

where $s$, $b$ and $SNIa$ are our yields from single stars, binary stars 
without SNeIa, and SNeIa yields only, respectively 
(see section~\ref{yield-calculations}). With these yield sets we simulated 
the chemical evolution of galaxies with various star formation histories 
(Table~\ref{SFH-table}) to compare the effect of binary star 
yields \emph{other than SNeIa} on line strengths. Our GCE code was run 
with each of the above two combinations of 
stellar populations (given in Equations 1 and 2), for the different SFHs. 
Differences between the two assumed populations were thus found for gas 
properties and stellar line-strengths.

\subsection{Line strength changes relative to observational errors}
\label{obs-errors}

The relative differences in line strengths between populations S and B, as a 
percentage of typical observational errors, are shown in Fig.~\ref{Fig-1}.
Table~\ref{Table-errors} shows the adopted errors which are described in 
detail in  Appendix~\ref{Line-strength-observational-errors}.

All five SFHs lead to similar differences in line-strength indices between 
populations B and S. The differences were less than $0.9\times$ typical 
observational errors, or less than $4\%$ of the index strengths themselves 
(for indices that remain positive). 

If we assume $\afe=0$ in our GCE model then the differences between populations
B and S are smaller, less than $0.75\times$ the typical error.
This reflects the fact that the metallicity and the $\afe$ ratio are
dominated by SNeIa and that the overall metallicity changes very little 
between GCE models which use yields B (Equation~\ref{Eq-1}) or S 
(Equation~\ref{Eq-2}).

\begin{table}
\begin{centering}
\begin{tabular}{cccccccc}
\hline

{Index}&{PS02} & {S-B06} & {Dea05$\times0.75$} \cr
\hline
H$\delta_A$ (\AA) & 0.561 & {\bf 0.169} & 0.188 \cr
H$\delta_F$ (\AA) & 0.156 & {\bf 0.107} & 0.130 \cr
CN$_1$ (mag)      & 0.0337 & {\bf 0.0054} & 0.0061 \cr
CN$_2$ (mag)      & 0.0315 & {\bf 0.0066} & 0.0068 \cr
Ca4227 (\AA)      & 0.071 & {\bf 0.076} & 0.087 \cr
G4300 (\AA)       & 0.559 & {\bf 0.185} & 0.123 \cr
H$\gamma_A$ (\AA) & 0.934 & {\bf 0.190} & 0.177 \cr
H$\gamma_F$ (\AA) & 0.859 & {\bf 0.108} & 0.106 \cr
Fe4383 (\AA)      & 1.715 & {\bf 0.208} & 0.170 \cr
Ca4455 (\AA)      & 0.401 & {\bf 0.093} & 0.083 \cr
Fe4531 (\AA)      & 0.590 & {\bf 0.131} & 0.141 \cr
C$_2$4668 (\AA)   & {\bf 0.173} & 0.253 & 0.208 \cr
H$\beta$ (\AA)    & {\bf 0.081} & 0.079 & 0.094 \cr
Fe5015 (\AA)      & {\bf 0.181} & 0.189 & 0.168 \cr
Mg$_1$ (mag)      & {\bf 0.0032} & - & 0.0024 \cr
Mg$_2$ (mag)      & {\bf 0.0041} & - & 0.0050 \cr
Mg$_b$ (\AA)      & {\bf 0.076} & 0.143 & 0.077 \cr
Fe5270 (\AA)      & {\bf 0.082} & - & 0.083 \cr
Fe5335 (\AA)      & {\bf 0.100} & - & 0.095 \cr
Fe5406 (\AA)      & {\bf 0.070} & - & 0.066 \cr
Fe5709 (\AA)      & - & - & {\bf 0.054} \cr
Fe5782 (\AA)      & - & - & {\bf 0.059} \cr
NaD (\AA)         & - & - & {\bf 0.070} \cr
TiO$_1$ (mag)     & - & - & {\bf 0.0018} \cr
TiO$_2$ (mag)     & - & - & {\bf 0.0020} \cr
&&&
 \cr 
\hline
 \end{tabular}
\caption{\label{Table-errors}Observational line-strength errors and 
their sources: PS02 = \citealp{Proctor2002}, S-B06 = \citealp{Sanchez2006},
Dea05 = \citealp{Denicolo2005}. Our selected errors are in bold, as used in 
Fig.~1 and described in Appendix~\ref{Line-strength-observational-errors}.
}

\end{centering}
\end{table}

 \vspace{0.3cm}

Line strengths are essentially pseudo-equivalent widths because a true 
continuum cannot be defined in complex galaxy spectra.
For a given spectral feature, a pseudo-continuum is defined by intensity 
levels within specified sidebands \citep[e.g.][]{Worthey1994a} which 
means a given line strength can be positive or negative  
(e.g. $\mathrm{H}\gamma$ and $\mathrm{H}\delta$).
For this reason the differences in line strengths presented in 
Figure~\ref{Fig-1} are normalised to typical observational errors. 
Figure~\ref{Fig-1} shows that the 
differences between line strengths calculated with populations using 
the yield sets B and S are all well within typical observational errors 
for all the SFHs tested. 
Fig.~\ref{Fig-2} illustrates these differences in terms of fractional 
changes, for those absorption line-strength that take only positive values. 

We note that some small, systematic effects are present in the line-strength
differences between B and S populations. Line strengths calculated
for yield set B typically have slightly smaller metal-sensitive features 
(coloured blue, orange and black in Fig.~\ref{Fig-1}) and slightly 
elevated hydrogen line strengths (coloured magenta in Fig.~\ref{Fig-1}). 
Calcium (in black, labelled Ca) and iron sensitive features (in blue, 
labelled Fe) are depreciated 
by typically $0.2\times$ the line strength errors.
Magnesium and carbon sensitive features 
(in orange and black respectively) show the most negative differences 
which implies that $\alpha$-capture metal absorption features are 
weaker in our models that include binary yields. 
These changes are a result of enhanced mass loss in binaries which leads
to ejection of material which has been less processed by nuclear burning
than in equivalent single stars.

In future, and with the advent of higher signal-to-noise data (by at 
least a factor of 3),
these systematic effects might become detectable, as small offsets to best 
fit population models. However, other model uncertainties currently 
outweigh this, such as the yields from massive star models 
\citep[e.g.][]{Woosley1995,Hirschi2005,Eldridge2008} and the 
still-unknown progenitors, and associated yields, 
of SNeIa \citep{Iwamoto1999,Yungelson2004,Tout2005}. 
Similarly, uncertainties in calibrations to the Lick standard system 
and in SSP models limit our ability to probe such small effects in real 
populations. More accurate line-strength systems, based on well calibrated 
stellar libraries \citep[e.g.][]{Sanchez2006a}, 
will allow line strengths to be measured more accurately and might 
help us to see such subtle differences in the future.

\subsection{Time dependent abundances}

\label{Time-dependent-abundances}

Fig.~\ref{Fig-3} shows the differences between ISM abundances for populations 
with (B) and without (S) the binary yields other than SNeIa, for the five 
GCE models from Table~\ref{SFH-table}, as a function of time.
Similar levels of differences are seen in all the SFHs modelled with the 
GCE code. Abundance differences are small (less than $7\, \%$ in $Z$, Mg 
and Fe and less than $25\,\%$ in C and N) and follow the systematic 
differences reflected in the line strengths.
That is, iron, magnesium, carbon, nitrogen and the total metallicity are 
typically slightly lower in abundance in the ISM of the population including 
other binary yields. Again this is because of binary-enhanced mass-loss of, 
relative to single-stars, relatively unprocessed material. 

\begin{figure*}
\begin{minipage}{150mm}
\includegraphics[angle=0,width=170mm,trim=0 0 0 0]{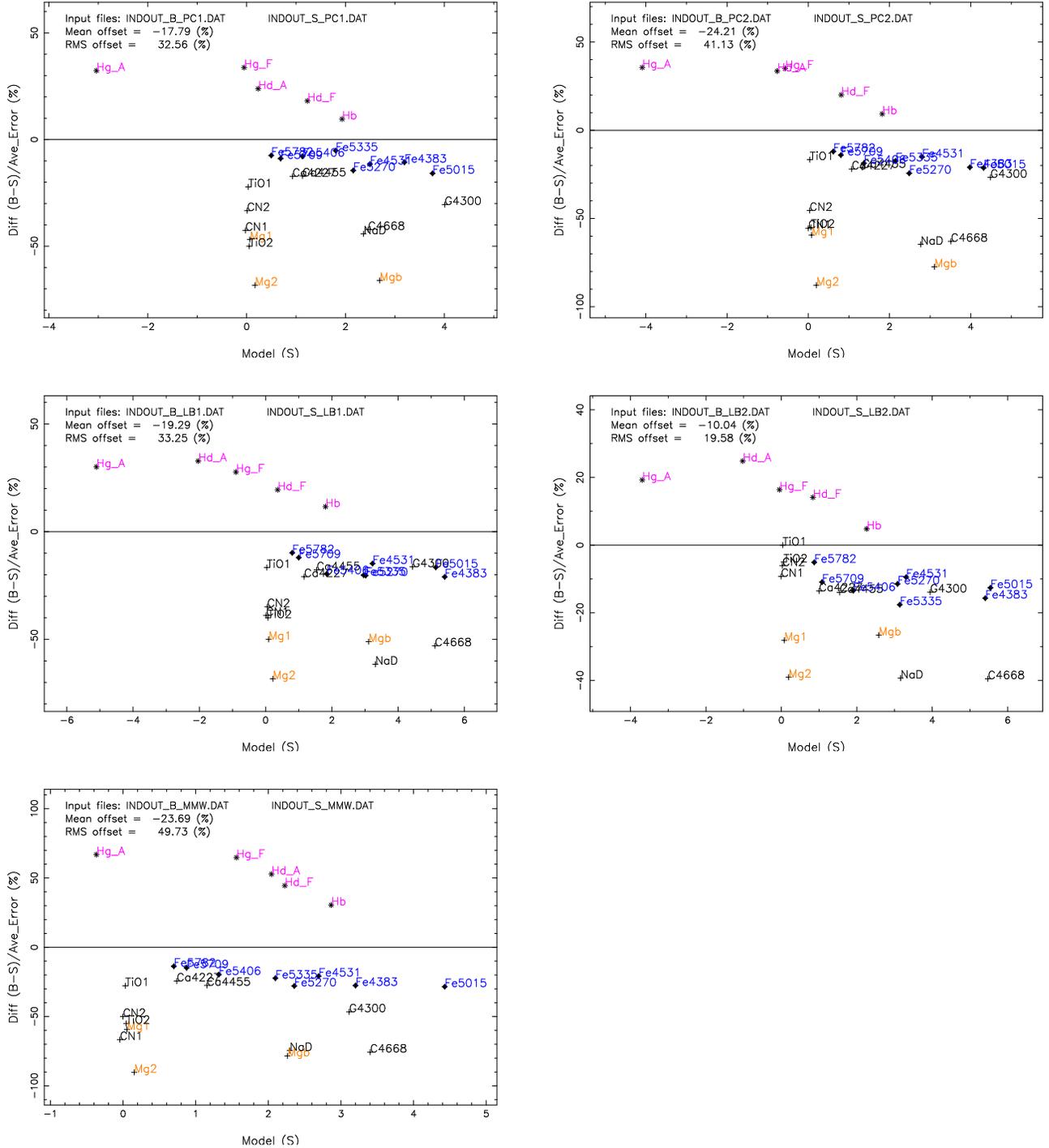}
 \caption{\label{Fig-1}Differences in indices found for populations 
with-without other binary yields (apart from SNeIa explosions), for the 
five different SFHs, plotted in the order given in Table~\ref{SFH-table}. 
The differences are plotted as a percentage of typical observational 
errors. The indices plotted include the line strengths and molecular band 
features listed in Table~\ref{Table-errors}. Hydrogen features are labelled 
in magenta (starred points), iron sensitive features are in blue 
(filled diamond points), magnesium sensitive features are in orange 
(pluses) and other features are in black (pluses).}
\end{minipage}
\end{figure*}

\begin{figure*}
\begin{minipage}{150mm}
\includegraphics[angle=0,width=170mm,trim=0 0 0 0]{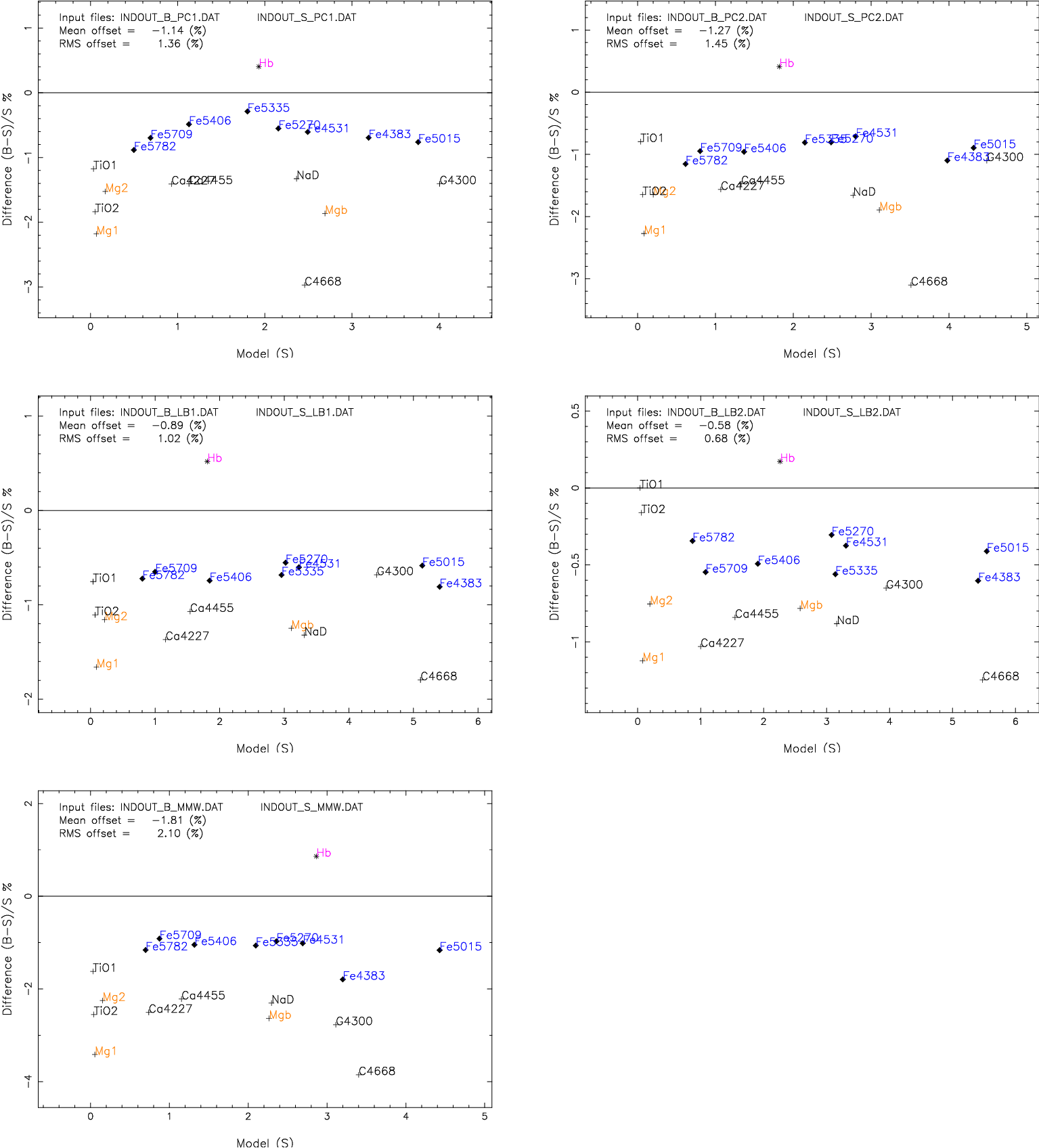}
 \caption{\label{Fig-2} Same as for Fig.~\ref{Fig-1}, except that here the
fractional differences are plotted for 
those line-strengths that take only positive values (i.e. excluding
$\mathrm{H}\gamma$, $\mathrm{H}\delta$ and CN bands).}
\end{minipage}
\end{figure*}

\begin{figure*}
\begin{minipage}{150mm}
\includegraphics[angle=0,width=170mm,trim=0 0 0 0]{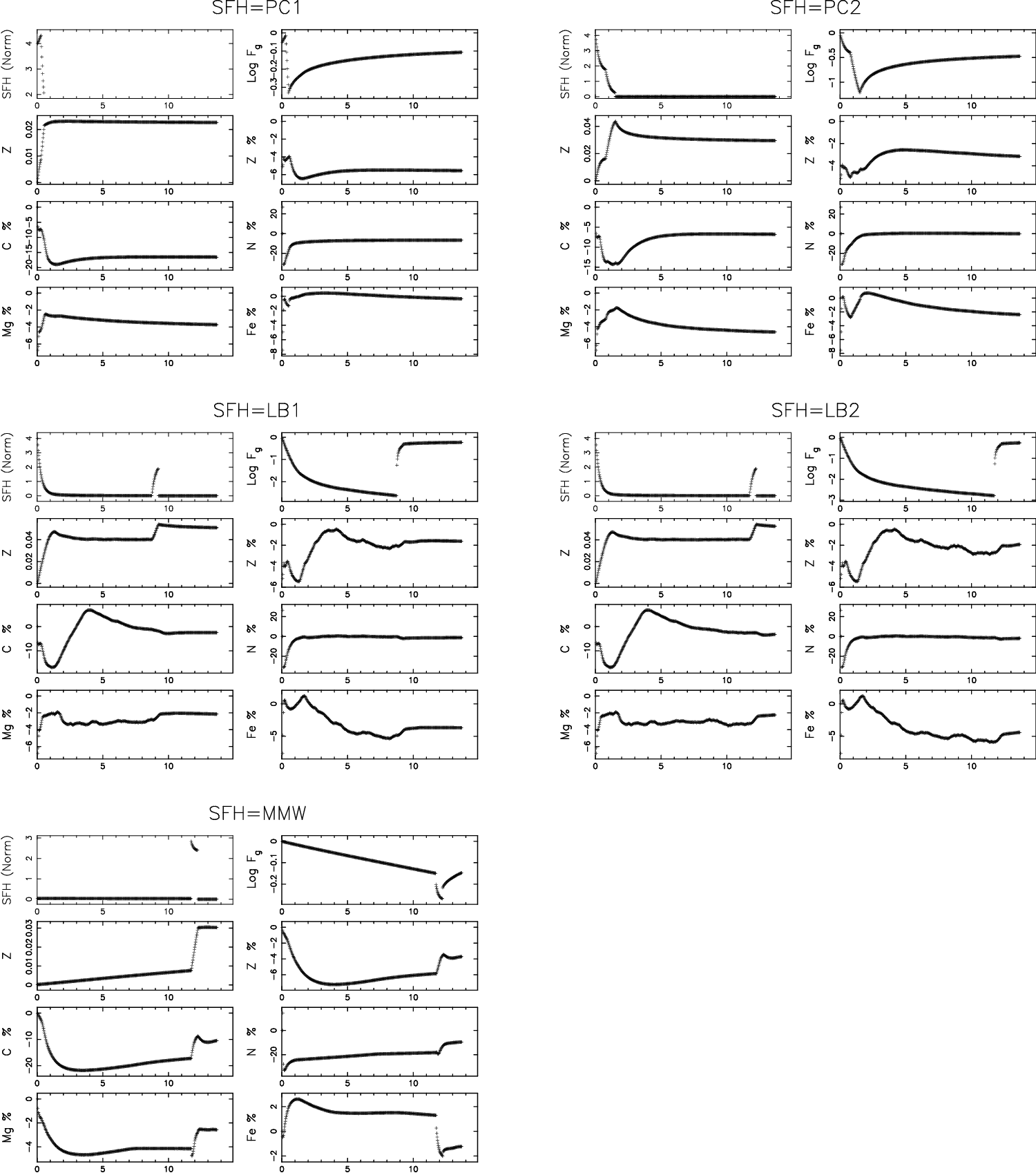}
\vspace{14pt}
 \caption{\label{Fig-3}Normalised SFH, normalised gas mass, total metallicity 
in the ISM and differences in ISM abundances for various elements as a 
function of time in Gyr, for the five star formation history models
listed in Table~\ref{SFH-table}.
Upper row shows results for SFH models PC1 (left) and PC2 (right), 
central row shows LB1 (left) and LB2 (right) models and lower row shows 
MMW model.
The SFH axis is the rate of mass going into stars per Gyr, normalised 
by the initial mass of gas. The gas mass fraction 
($\mathrm{F}_g$) is also normalised by the initial mass of gas.
Total metallicities are in mass fractions and abundance differences 
are in percentage, i.e. 100(B-S)/B, where B and S are the populations 
with and without other binaries respectively (see Equations 1 and 2). 
N.B. The oscillations in our models with late bursts (LB1 and LB2) 
are a result of low gas masses at those times.}
\end{minipage}
\end{figure*}

Oscillations in ISM gas abundances due to small numbers of stars in 
the yield calculations can be seen when the gas mass and 
feedback from older stars are small. The GCE model assumes a gas inflow 
composition the same as the existing ISM which aggravates the problem. 
For example gas inflow in the late burst (LB) models only starts at T1, 
so the composition of the ISM then strongly influences the subsequent 
chemistry in our models, even if there was hardly any ISM left just 
prior to T1. This problem is drastically reduced (but not totally eliminated) 
by using sufficient numbers of stars in the evaluations of population yields
and fine time steps in those evaluations when the evolution is most rapid
(see e.g. Fig.~\ref{Fig-2} where the LB models behave overall in a similar 
way as the other models plotted there, and Fig.~\ref{Fig-3} where the 
oscillations in yields are relatively small).
The problem does not affect the vast majority of stars in our simulated 
galaxies because the rate of star formation is proportional to the gas mass.

\section{Different SNeIa}
\label{section-different-sneia}
The contributions of Fe peak elements from SNeIa enrichment are crucial to 
the correct understanding and interpretation of galaxy spectra through 
GCE modelling. Both the timescale and quantity of the enrichment pattern
from SNeIa are important. Most modellers of stellar populations (integrated
or resolved) ignore the uncertainties in SNeIa models and use the 
predictions of yields from the standard COWD deflagration model for 
SNeIa (W7 model of \citealp*{Nomoto1984}). Detailed binary-star models 
(e.g. \citealp{DeDonder2004}; \citealp{Han2004}; \citealp[][and references 
therein]{Tout2005}; \citealp*{Martin2006}) indicate 
how difficult it is to form SNeIa through this conventionally assumed route 
of a COWD increasing its mass through steady accretion and burning of 
material from an interacting companion star until it reaches the 
Chandrasekhar mass limit ($\Mch \simeq 1.4\,\,\msun$). The difficulty 
arises in the unstable nature of the accretion: too rapid and a common 
envelope may form, which does not necessarily lead to a SNIa; too slow 
and novae (probably) lead to mass loss.
The theoretical mass accretion rates under which a
standard SNIa can form cover only a factor of a few up to a few 
$\times 10^{-7}\,\msun\,\mathrm{yr}^{-1}$ (\citealp{Tout2005} 
and references therein). 

The results of \citet*{Hachisu1996} and \citet*{Hachisu2008} suggest 
that accretion rates of 
$\sim 10^{-7}$ to $10^{-5}\msunpy$ could lead 
to SNeIa if the accreting white dwarf has a strong wind that effectively 
reduces the accretion rate to that required for steady accretion.
Here we refer to this as the 'disk wind' model.

There are other channels that may lead to a Chandrasekhar-mass COWD and 
hence a Ia supernova. These include close double-degenerate (COWD-COWD) 
systems, which merge because of orbital shrinkage caused by gravitational 
radiation. Theoretically, He white dwarfs in binaries may also contribute 
to SNIa, however, evidence for their existence as SNeIa progenitors is 
lacking (\citealp{Tout2005}).

Sub-Chandrasekhar mass models, while currently out of favour as Ia 
progenitors, dominate the rate of SNeIa in our binary models. These are 
mostly edge-lit detonations (ELDs) of COWDs accreting $0.15\msun$ of 
helium-rich material which ignites, setting off the whole star.
If these types of SNeIa really do dominate the yields then it is important 
to study whether there are any chemical signatures that could be used to 
confirm their presence in integrated stellar populations.

Table~\ref{Table-SNIa-yields} shows the relative contributions from different 
types of SNeIa from our binary yields at different metallicities assuming our 
baseline binary-star plus single-star model (Section~\ref{yield-calculations}).
This shows that yields from ELDs dominate the SNIa contributions in our 
models, particularly at high metallicity. 
The exact mix of SNIa types does not matter for the tests of the impact 
of other binary yields (from common-envelope evolution, novae etc.) 
on line strengths (Section~\ref{obs-errors} above), but it may limit 
our ability to accurately interpret line strengths from galaxies. Therefore
we go on to try to test the impact of different SNeIa on spectral 
line strengths. 

\begin{figure*}
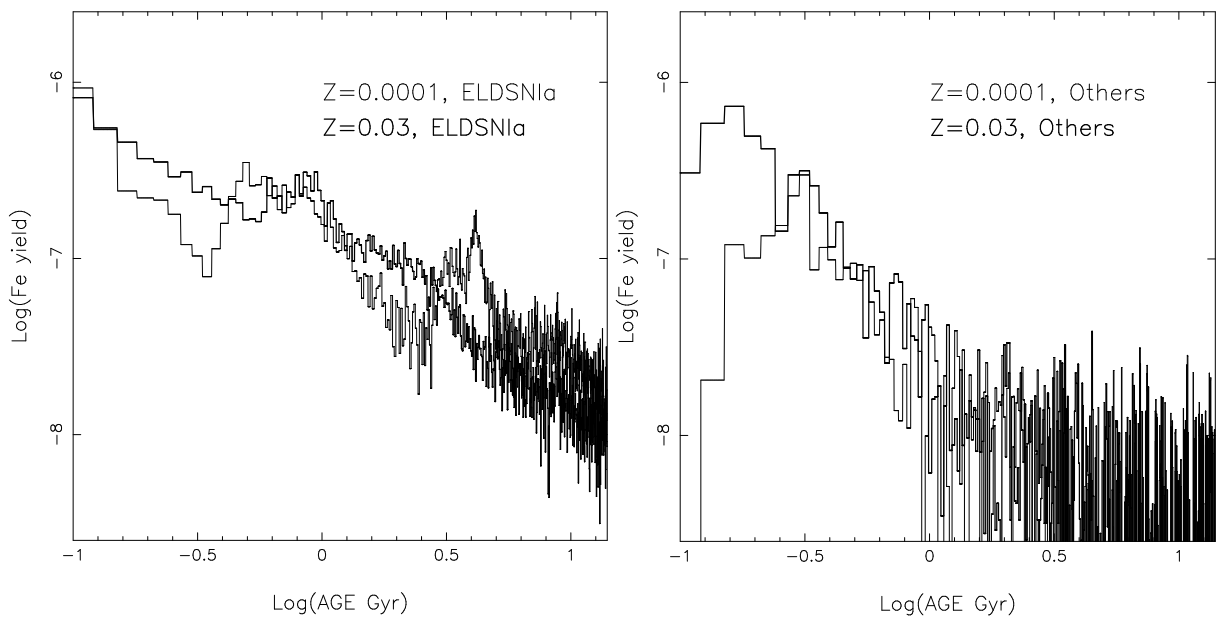

\begin{minipage}{150mm}
\includegraphics[angle=-90,width=80mm,trim=0 0 0 0]{Fe_ELDSNIa.ps}
\includegraphics[angle=-90,width=80mm,trim=0 0 0 0]{Fe_Others.ps}
\caption{\label{Fig-4}Time evolution of iron yields from two different 
groups of SNeIa, from binary star model yields. The models shown are 
for ELD and Other SNeIa as listed in Table~\ref{Table-SNIa-yields}.}
\end{minipage}
\end{figure*}

As expected, iron dominates the yields from the SNeIa in these models 
(see Table~\ref{Table-SNIa-yields}). 
Fig.~\ref{Fig-4} shows the time evolution of iron for two different 
categories of SNeIa. Note from these plots 
how the distributions are sensitive to the initial metallicity of the stars.
For all the possible SNeIa model yields shown in Fig.~\ref{Fig-4} the 
feedback from SNeIa includes a large peak at very early times 
($\sim 10^8$ yr) plus extended feedback at a lower level, generally 
decreasing over the 
rest of the time. This is in contrast to what was often assumed until 
recently, where the SNeIa contributions were not expected to contribute 
significantly in galaxies until a few $\times 10^8$ yr 
(\citealp{Matteucci2001}). It is, however, much more in agreement with 
recent ideas about SNeIa timescales (e.g. \citealp{Aubourg2008}; 
\citealp{Hachisu2008} and references therein; \citealp{Kobayashi2008}), 
in which SNIa occur with short and long delay times.

\begin{table*}
\begin{minipage}{150mm}
\begin{centering}
\begin{tabular}{l r r r r r r r}
\hline
{\bf{$Z$}}&0.0001& 0.0005& 0.001& 0.005& 0.01& 0.02& 0.03 \cr 
\hline
{\bf{Total}}&0.0061& 0.0058& 0.0051& 0.0035& 0.0032& 0.0031& 0.0029 \cr
{\bf{Fe}}   &0.0038& 0.0036& 0.0032& 0.0019& 0.0016& 0.0015& 0.0014 \cr
{\bf{ELD} (\%)}&30.1& 29.1& 30.0& 53.9& 60.8& 63.3& 65.8 \cr 
{\bf{HeIa} (\%)}& 66.6& 64.0& 63.2& 37.1& 28.3& 24.9& 21.8 \cr 
{\bf{Others} (\%)}& 3.3& 6.8& 6.8& 9.0& 10.9& 11.8& 12.4 \cr
\hline
 \end{tabular}
\caption{\label{Table-SNIa-yields}Mass contributions of various types 
of SNIa progenitors to the total SNIa yields as a function of 
metallicity $Z$ for our baseline model. The ``Total'' is 
the mass fraction returned from SNeIa relative to the total initial 
mass of the stellar population.
Similarly, the mass fraction returned from iron is shown as ``Fe''. 
The final three rows show percentages of ejecta from edge-lit 
detonations (sub-$\Mch$); He white dwarf explosions and
other ($\Mch$) explosions e.g. accreting COWD and COWD-COWD mergers.
} 
\end{centering}
\end{minipage}
\end{table*}

 \vspace{0.3cm}

\begin{table*}

\begin{minipage}{150mm}
\begin{centering}
\hspace{-1cm}  
\begin{tabular}{llccccccccc}
\hline

\bf{Test} &\bf{Scenario} & \multicolumn{5}{c}{\bf Proportions}  &\bf{Max diff} &\bf{RMS} & \bf{Comment} \cr
     
\bf{Number}  &  &\bf{s}   & \bf{b}   & \bf{All SNIa}   & \bf{ELD} &\bf{Others} &\bf{/errors} &\bf{diff} & \cr

& & & &  & \bf{only} & \bf{only}  & & & \cr

\hline
1 &Baseline                 &0.5 &0.5 &0.5   &0   &0     & 0    & 0    & (B, Eq.~1) \cr
2 &Single stars + SNIa      &1.0 &0   &0.5   &0   &0     & 0.9 & 0.50 & (S in Eq.~2) \cr
3 &$\ace=1.0$               &0.5 &0.5 &0.5   &0   &0     & 1.0  & 0.39 &  \cr
4 &$\lce=0.5$               &0.5 &0.5 &0.5   &0   &0     & 0.65 & 0.30 &  \cr
5 &Standard SNIa            &0.5 &0.5 &0     &0   &4.237 & 1.0  & 0.46 &  $\Mch$ COWDs \cr
6 &Exclude all SNIa         &0.5 &0.5 &0     &0   &0     & 3.2  & 1.71 &  \cr
7 &$2\times$ baseline SNIa  &0.5 &0.5 &1.0   &0   &0     & 2.0  & 0.95 &  \cr
8 &Disk wind                &0.5 &0.5 &0.5   &0   &0     & 4.0  & 1.97 &  \cr
&&&&&&&&&& \cr 
\hline
 \end{tabular}
\caption{\label{Table-SNIa-tests}Proportions used in tests and resultant 
differences in line-strengths compared to the baseline model, 
which includes other binary yields as well as those from SNeIa 
(see Section ~\ref{yield-calculations}).
Maximum and RMS differences are given, expressed as a fraction of 
the typical observational errors on line-strengths. 
From Section~\ref{yield-calculations}: s are single stars, b are binaries 
excluding SNeIa. Tests 3, 4 and 8 used different SNIa calculations. 
All differences are compared with the baseline model. 
See section~\ref{effects-of-different-sneia} for discussion 
of the sense of the differences for particular groups of line-strengths.}

\end{centering}
\end{minipage}
\end{table*}
 \vspace{0.3cm}
 
\subsection{The effect of different SNIa progenitors and 
common envelope parameters} 
\label{effects-of-different-sneia}
In this section we carry out a series of tests to examine the effects on 
line strengths from different combinations of SNIa types and common 
envelope parameters using each the SFHs given in Table~\ref{SFH-table}.
Our GCE model was modified to allow yields from different proportions 
of the various SNIa types. Yield sets were calculated from weighted 
sums of the following subsets of the baseline yields: \\ 
 
\renewcommand{\labelenumiv}{\roman{enumiv}}
\begin{enumerate}
\item\setlength{\parskip}{-10pt} Single stars, \\ 
\item Binary stars (excluding SNeIa), \\  
\item SNeIa (baseline model), \\ 
\item He WD SNeIa (accreting and merging), \\ 
\item ELD SNeIa, \\ 
\item Others (COWDs which accrete or merge such that  $M_{\rm{WD}}>\Mch$) 
\end{enumerate}

where (iii)$=$(iv)$+$(v)$+$(vi).

That is, the sum of equal weight contributions from iv, v and vi above 
add up to the baseline model SNIa contributions (iii above). 
Option vi labelled 'Others'
consists of yields from Chandrasekhar-mass SNIa explosions. 

Table~\ref{Table-SNIa-tests} lists the eight test cases with corresponding 
weighting factors. The results of each test are then compared with the 
baseline model (which uses proportions given in Equation~\ref{Eq-1}, 
equivalent to test~1). Differences that arise are then due to the 
proportions of SNIa types included. The maximum difference in any line 
strength, relative to the relevant observational error and the root mean 
square difference in all line strengths relative to observation errors 
are also shown in Table~\ref{Table-SNIa-tests}. 

Tests 1 and 2 are equivalent to our populations B, the ``baseline model'', 
and S (Eqs.~\ref{Eq-1} and \ref{Eq-2} respectively). Again this shows that 
the effect of changing from a purely single-star population with SNeIa to 
a true half-single and half-binary population is within the observational 
uncertainties on the line strengths.

Parameters that quantify the behaviour of 
a common envelope (CE) formed around a close binary can affect the 
number and type of SNeIa that occur, particularly the number of double 
degenerates that can merge to form SNeIa. 
The efficiency of orbital energy transfer to the CE is parameterised by 
$\ace$ (see Hurley et al. 2002). The CE may be completely removed 
by orbital energy transfer to the envelope. Observational evidence for 
this is in the existence of close double degenerate binaries such as 
4U1820-30 or B2303+46 (Church et al. 2006) which are thought to have 
achieved their close orbits via previous orbital energy loss to a CE.
Estimates of $\ace$ vary from less than 1 up to about 3, with $\ace>1$ 
possible if energy sources other than orbital energy are involved 
(Hurley et al. 2002).
If $\ace$ is large then energy is more easily transferred to the CE 
until it becomes unbound. Conversely, if the transfer of orbital energy 
to the CE is less efficient, then the CE may be bound for longer, giving 
the stellar orbits time to decay by various means, which is more likely 
to lead to coalescence (Church et al. 2006). However, if orbital energy 
transfer is too inefficient this essentially removes one of the potential 
orbital decay mechanisms for the binary, leading to fewer SNeIa via 
coalescence.

Another parameter relating to CE evolution is $\lce$, a dimensionless 
factor inversely proportional to the initial binding energy of the CE 
(\citet*{Hurley2002}). This parameter is important since it affects the 
amount of energy that can be transferred to the CE before it becomes unbound.
This affects the closeness of the two stellar cores remaining after the CE 
is lost. A value of $\lce=0.5$ was used in Hurley et al. and we 
test this value compared to our baseline model described in 
Section~\ref{yield-calculations}.

Tests 3 and 4 show that the effect on line strengths of changing the 
common envelope parameters by setting either $\ace=1$ or $\lce=0.5$ is 
about one third of the observational uncertainty and at most of the 
same order as the observational uncertainty. 
Therefore uncertainties in the values of common-envelope parameters, 
and their effects on SNeIa yields, can normally be neglected in GCE models 
of line-strengths in integrated stellar populations.

We now ask if there is a significant difference between using the baseline 
model yields (dominated by sub-$\Mch$ ELDs) and yields from $\Mch$ SNeIa 
scaled up to eject the same total mass (at $Z=0.02$). The results are 
shown in test~5. 
The $\Mch$ yields differ in both composition and rate from the 
sub-$\Mch$ yields, yet the maximum difference between the line 
strengths in both models was only of the order of the observational 
uncertainties with RMS differences of only half the observational 
uncertainty. Therefore it is difficult to distinguish between $\Mch$ 
and sub-$\Mch$ SNeIa from integrated spectral line strengths. 
However, the timing of SNeIa is 
important and our current models indicate that SNIa yields may 
contribute to galactic chemical evolution earlier than expected.    
 
Test~6 shows that the effect of removing SNeIa altogether is rather large, 
up to $3.2$ $\times$ the typical observational errors. Clearly this is 
detectable.
The primordial collapse models (PC1 and PC2) showed smaller differences 
because in those cases SNeIa do not have much time to make an impact 
before star formation stops. This test illustrates the importance of including
SNIa yields to correctly interpret observed line strengths in galaxies.

Observed SNIa rates in galaxies cover about a factor of two in range
\citep[][their Table~5]{Navasardyan2001}. 
Models with double the baseline SNIa contribution were generated with
proportions given by test~7 in Table~\ref{Table-SNIa-tests}.
Differences were found to be up to twice the typical observational 
uncertainties so in principle may be detectable for some lines, 
such as iron-sensitive lines. Smaller differences were again found for 
the primordial models.
The iron sensitive line-strengths are slightly stronger for the doubled 
SNIa case compared with the baseline case (up to 4\% of the line-strengths 
themselves, and not very sensitive to the exact SFH). 
This is in the sense expected, since SNeIa contribute iron to the cosmic 
cycle.

Test~8 shows the effect of including a disk wind to stabilise accretion. 
In our models, this is achieved by allowing high mass accretion rates to 
lead to a SNIa, in order to test possible SNIa routes first described 
by \citet*{Hachisu1996} and subsequent work by those authors. 
As in tests 6 and 7, this shows that the line strengths are most strongly 
affected by the rate of SNeIa. The largest differences were found in 
models with late bursts, as expected from the delayed SNIa contributions 
in these models.

\section{Discussion}
\label{section-discussion}
We have shown, using our chemical evolution models, that varying parameters 
that affect the number and type of SNeIa has a greater impact on line 
strengths than inclusion or exclusion of yields from other binary stars. 
This highlights the need to focus on the effects of different SNIa models 
in GCE codes. The effects of other binary yields (including common-envelope 
systems that do not become SNIa, CVs, novae etc) can largely be ignored, 
unless specific, low abundance isotopes, 
such as $^{13}$C or $^{15}$N are being studied \citep{Izzard2006}. For 
most GCE applications, where abundances of Fe, Mg, C, Ca and overall 
metallicity have the largest impact on line strengths, it is not 
so necessary to include additional binary yields other than from SNeIa 
explosions.

The effects of additional binary yields other than SNeIa explosions 
are smaller than the observational uncertainties on line-strengths.
This answers our original question in that we do not find that it 
is necessary to include effects of other binary processes besides 
SNeIa explosions in GCE models of integrated stellar populations in galaxies. 
This may change if a factor of about three improvement in signal-to-noise 
occurs in line-strengths, based on a new, more accurate calibration system 
than the Lick star data.

We can see from Table~\ref{Table-SNIa-tests} that overpopulating with SNeIa
(test~8) or excluding SNeIa altogether (test~6) makes the biggest 
difference to predicted line strengths and these are clearly unrealistic 
and testable. After that, the rate and type of SNeIa are important. 
The adopted treatment of common envelope evolution makes a small difference, 
but less than the observational uncertainties in line strengths.

\citet{Matteucci2006} tested the effects of a more
bimodal SNIa rate distribution, using analytical forms, including prompt 
SNeIa ($<10^8$ yr) and tardy SNeIa (up to 10 Gyr) components. They predicted 
SNIa rates and yields for specific SFHs. They found, for ellipticals
modelled as early, primordial bursts, that the exact SNIa rate distribution 
had little effect on SNIa rates at late times. Our binary-star yields do 
include a prompt SNIa ($<10^8$ yr) and a tardy (continuing) SNIa 
contribution for both ELDs and $\Mch$ SNIa types (see Figure~\ref{Fig-4}).
Observational evidence for short ($<$70 Myr) and long (few Gyr) SNIa delay 
times is given by \citet{Aubourg2008}, who look at delay timescales of 
SNeIa following star formation epochs, in a large sample of galaxies.
By contrast, \citet{Forster2006} find that no progenitor model can be
ruled in or out with current data. 

We find that He WD progenitors for SNeIa may have a small effect in GCE 
models, however there is as yet no observational evidence for these, 
therefore they are largely ignored in this work. If real, their timescales 
for contributions may be longer than other SNeIa, due to the low mass 
progenitors involved.

Forcing the baseline SNIa yields to originate from $\Mch$ COWDs only, 
by only including scaled-up versions of those contributions led to relatively 
small differences in line strengths compared to our baseline model. 
Therefore, whether we model SNeIa as ELDs or standard $\Mch$ 
COWDs makes surprisingly little difference to predicted line-strengths 
in the models now. However, the yields do show larger variations with 
time, which might be detectable in resolved stellar populations.
Thus for the SNIa explosions themselves, GCE modelling of integrated 
populations does not appear to distinguish well between ELDs and 
standard SNeIa.

\subsection{Effects on average age and composition estimates} 
\label{effects-on-averages}
Finally, we consider what the implications are for single SSP fitting.

Fitting one SSP to spectral line strengths in order to recover luminosity 
weighted average ages and compositions is now a widely used technique 
in galaxy evolution studies. This usually involves a look-up table of SSPs 
on a grid of values in age, metallicity and sometimes abundance ratio 
parameter space. The grid spacing we use in our SSP fitting routines 
(based on TMK04 SSPs and code written by Robert Proctor) is 
0.025 in Log(Age) and Log($Z$), with coarser grid spacing for abundance 
ratio. This translates to $\sim 6\, \%$ differences between 
adjacent grid points in age and metallicity. Thus the 2\% to 7\% lower $Z$ in 
composite populations including other binaries (seen in Fig.~\ref{Fig-3},
$Z$\% panels) gives rise to one grid spacing lower in luminosity weighted 
average metallicity ($Z$). 

For luminosity weighted average ages we find no difference between estimates
from line strengths with or without other binaries. The time of the latest 
starburst (see Fig.~\ref{Fig-3}, SFH panels) is approximately recovered in 
most cases, with no systematic differences for fits to line strength with or 
without other binaries. In the case of the model PC2, the average age is 
estimated at 8.4 Gyr in each case, whereas the stars in that model really 
formed between 13.7 and 12.2 Gyr ago. This highlights the fact that SSP fitting
is far more strongly affected by aspects other than the effects of other 
binaries besides SNIa. The fact that average age estimates are not affected
by other binaries is a bit surprising, given that all the Balmer indices are
stronger when the yields from other binaries are included 
(see Fig.~\ref{Fig-1}). However, the stronger Balmer indices are offset by 
other age sensitive features which are systematically weaker (e.g. G4300) 
and it is important to remember that the Balmer indices do have some 
sensitivity to metallicity as well - for a given age, they are stronger 
at lower metallicity. 

Thus average age estimates are unaffected by yields from other binaries 
(such as common envelope systems, CVs and Novae) whilst average metallicities 
are slightly reduced (by $<7\,\%$) in populations accounting for yields from 
these other binaries. Differences in SNIa types are likely to produce larger
effects, but SNeIa are as yet too poorly understood to address this issue with
any accuracy. Future work to uncover the progenitors of SNeIa will help to
address this large uncertainty in galactic chemical evolution modelling.
From our present study we can say that other binaries aside from SNeIa have 
only a small effect on integrated spectra of galaxies and so do not normally 
need to be incorporated to interpret galaxy ages and metallicities.

\section{Conclusions}
\label{section-conclusions}
In this paper we have investigated the effects of yields from binary stars
on spectral line strengths in integrated stellar populations. To do this we 
combined a single-/binary-star population model, which predicts 
yields as a function of time, with a galactic chemical 
evolution code that self-consistently models the spectral line strengths 
from integrated populations with different star formation histories.
Our tested SFHs included stars more than $1.5$Gyr old up to 13.7Gyr 
(Table~\ref{SFH-table}).
Derived line strengths for various assumed binary-star 
contributions were compared to a baseline model. The resulting differences 
were compared to typical observational uncertainties on line strengths, 
in order to illustrate the relative importance of the various contributions 
and binary effects. We find the following:

\begin{itemize}
\item Populations including other binary star processes as well as 
SNeIa produce 
slightly less metal than a population consisting of single stars with SNeIa.
This affects magnesium and carbon sensitive spectral features most and 
iron sensitive features least (Figs.~\ref{Fig-1} and \ref{Fig-2}). 
Hydrogen absorption lines are slightly enhanced in 
populations with other binaries. However, the differences are all less 
than $0.9 \times$ typical observational uncertainties on line-strengths.
\item Therefore galactic chemical evolution models of unresolved stellar 
populations, using Lick spectral indices, do not need to incorporate 
yields from 
binary stars other than SNIa explosions because other binary processes
(such as CVs, novae, symbiotic stars and various common envelope binaries) 
do not significantly alter the strengths of observed spectral features.
\item Parameters describing common envelope evolution affect the number of 
SNeIa. Varying these parameters (e.g. efficiency of orbital energy transfer 
to the CE) affected observed line strengths by up to the observational errors 
in the current simulations. Therefore these effects are also quite small 
(less than or equal to observational errors). 
\item The exact nature of the explosion (e.g. Sub-$\Mch$ or $\Mch$ COWD) makes 
less difference than the rate of SNeIa and the time variation of that rate.
\item Reductions in observational errors by about a factor of three are 
needed for line strengths to be sensitive to yields from other binary 
stars besides SNeIa, or the effects of CE evolution, within the bounds 
of currently acceptable parameters. On the other hand, current measurements 
of line strengths are
already sensitive to the rates and timescales of SNeIa. However, they are 
also at least as sensitive to uncertainties in massive star evolution.
\item Luminosity weighted average ages are unaffected by yields from other
binaries and luminosity weighted average metallicities are decreased by 
less than $7\,\%$ due to yields from other binaries.
\end{itemize}

This study makes a preliminary investigation into the
expected contributions of different SNIa types to
the chemical evolution of galaxies. The different timescales of 
SNIa versus SNII enrichment forms a vital argument in our understanding of
galaxy evolution from composite stellar populations. Thus, understanding the
evolution, timescales and chemical contributions of different types of 
SNeIa is fundamental for accurate GCE modelling. Therefore it is important 
in future work to further explore the accuracy and effects of SNIa 
models and their relative contributions in galaxies. Our theoretical 
yields indicate prompt ($\le 10^8$yr) as well as longer timescale 
contributions from SNeIa (Fig.~\ref{Fig-4}).

In this paper we have looked at the effects on spectral absorption 
line-strengths of varying assumptions about the contributions of yields
from different types of binary stars, in chemical evolution models of 
galaxies. In future work we will continue to improve our yield models, 
guided by observational results (e.g. from SuperWasp, CoRoT, Gaia and 
other variability and radial velocity
surveys). As more detailed modelling of individual elements become 
available we will also look at the effects on line strength of 
individual heavy elements such as carbon, which shows the largest 
differences in line-strengths, due to inclusion of different 
binary star processes other than SNeIa.

\section*{Acknowledgments}

We thank the University of Central Lancashire for a Livesey award that 
provided visitor funding during this work and also for postdoctoral 
funding that supported PO for one year at UCLan. Thanks to Robert Proctor 
for the use of his SSP fitting software. Thanks to the referee Guy Worthey 
for helpful comments. A postgraduate student Kate Bird helped to write a 
subroutine to read in the TMK04 SSP models. RGI thanks the Nederlandse 
Organisatie voor Wetenschappelijk Onderzoek for funding and is the 
recipient of a Marie Curie-Intra European Fellowship.

\appendix

\section{GCE parameters}
\label{GCE-parameters}
Parameterisation of the SFH used in our current GCE models is aimed at 
covering as wide a range of types of composite galaxy histories as possible 
with a minimum number of free parameters. We also want to specify the 
latest major burst in some detail because more recent star formation has 
a large effect on the luminosity, colours and integrated spectral properties
of the composite stellar population. Since an SSP is typically 
described by three parameters (age, $\feh$, $\afe$) we need more 
than three parameters to describe a composite SFH. Empirical fitting of two 
SSPs typically takes six parameters. We aim to improve on that, to describe 
more realistic continuous SFHs. For a GCE model incorporating a SFH, we have
additional constraints of self-consistency, which arise naturally from the 
model. Therefore the free parameters are different from the SSP case. They
do not describe the attributes of the stellar population directly, but instead
describe the star formation efficiency and gas flow properties as a function 
of time. 

\begin{figure*}
\begin{minipage}{150mm}
\includegraphics[angle=-90,width=180mm,trim=0 0 0 0]{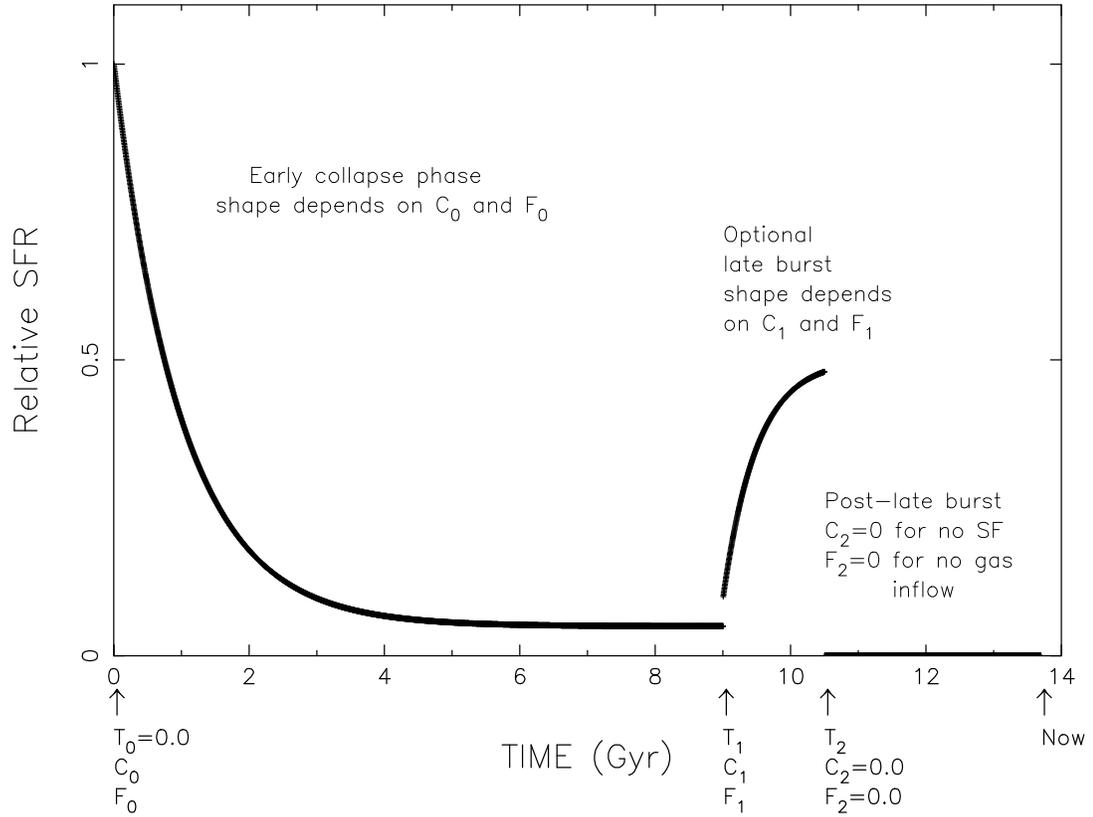}
\vspace{14pt}
 \caption{\label{SFH-schematic}Schematic illustration of the parameters 
used to specify a 
continuous star formation history in our galactic chemical evolution model.
The six free parameters (without values attached) 
and three fixed parameters (with values) are indicated below the time axis. 
In Table~\ref{SFH-table} values 
for the six free parameters are specified to describe different models tried 
in our analysis in this paper.}
\end{minipage}
\end{figure*}

The early star formation is parameterised by a star formation efficiency 
({\bf C$_0$}) and a gas flow rate ({\bf F$_0$}).
To describe early star formation, plus details of a late (optional)
burst, we allow for two interruptions to the history at times T$>0.0$. 
These times are represented by {\bf T$_1$} and T$_2$, as indicated in 
Fig.~\ref{SFH-schematic}. T$_1$ can occur anywhere in the range 
$0<$T$_1<12.2$ Gyr, i.e. 1.5 Gyr ago or older in the current work, 
assuming an age of the 
universe of 13.7 Gyr. The duration of the latest burst is then 
{\bf D$_1$}=T$_2$-T$_1$, as in Table~\ref{SFH-table}. Star formation 
efficiency ({\bf C$_1$}) and gas flow rate ({\bf F$_1$}) 
are both set as free parameters at T$_1$. Then at T$_2$ the star formation 
and gas flow are switched off (by fixing C$_2$=0 and F$_2$=0) because we are 
not attempting to model systems with recent or ongoing star formation.
The SSPs used in this current work are 1.5 Gyr or older, so this is 
the limit of how young we model stars. This gives us six parameters 
(the ones given in bold above) that can be varied to describe a wide 
range of histories much more realistically than two SSPs (which also 
requires 6 parameters as mentioned above).

Fig.~\ref{SFH-schematic} illustrates the SFH parameters described, 
along with a time 
sequence covering the age of the Universe. We are not sensitive to the 
exact age of the oldest stars because a population that is $\sim$10 Gyr old 
looks very similar to one that is $\sim$14 Gyr old in its spectrum and 
colours. However we are still sensitive to the burst duration in old 
populations, through abundance ratios. Utilizing the six parameters 
described above we can explore histories ranging from early collapse 
(such as models PC1 and PC2 in Table~\ref{SFH-table}), 
early star formation plus a later burst (such as models LB1, LB2 and 
MMW in Table~\ref{SFH-table}), through to continuous star formation 
ending recently.
The latter models are not explored here for modelling effects of binary 
stars in luminous early-type galaxies because $\afe$ ratios in these
galaxies are not consistent with such models 
(e.g. \citealp{Kuntschner2002}; PS02) and these 
galaxies do not generally contain significant ongoing star formation.

\section{Line strength observational uncertainties}
\label{Line-strength-observational-errors}
We estimate average uncertainties on line strengths from observations 
of early-type galaxies (PS02; \citealp{Denicolo2005} - Dea05 
in Table~\ref{Table-errors}; \citealp{Sanchez2006} - S-B06 
in Table~\ref{Table-errors}) acquired with $4\,\mathrm{m}$ class 
telescopes (e.g. 17 E/S0 galaxies from PS02, observed with 
the $4.2\,\mathrm{m}$ William Herschel telescope).
 
We include the data from PS02 for wavelengths above $4600 \mathrm{\AA}$, 
below this the data are affected by dichroic response uncertainties.
Shorter than $4600 \mathrm{\AA}$ we use the data of \citet{Sanchez2006}
who provide data on 98 early-type galaxies as observed with the 
William Herschel telescope or the $3.5\,\mathrm{m}$ Calar Alto telescope. 

At wavelengths longer than $5700 \mathrm{\AA}$ errors were obtained from 
observations of by \citet{Denicolo2005}.  They catalogue 93 early-type galaxies
observed with the $2.12\,\mathrm{m}$ telescope in Mexico, 
so averages of their errors are scaled by a factor $0.75$ 
(the ratio of line-strength errors from PS02 to Dea05 in the spectral 
range $4600$~to~$5500\mathrm{\AA}$). 
In this way, our line-strength errors are 
estimated from observations with the equivalent of 
a $\sim4\,\mathrm{m}$ class telescope 
and a typical exposure time of $1$ to $2$ hours. 
Deeper observations or the use of a larger telescope
will not drastically reduce these errors, because they are 
dominated by systematic errors in the calibration to 
the Lick scale rather than Poisson errors (e.g. see 
\citealp{Proctor2000}, their Table~2; PS02, their 
Table~2; \citealp{Sanchez2006}, their Table~4). 
Table~\ref{Table-errors} highlights the errors used in the 
normalisations for Fig.~\ref{Fig-1}.

\end{document}